\documentclass[conference]{IEEEtran}

\usepackage{cite}
\usepackage{amsmath,amssymb,amsfonts}
\usepackage[hidelinks]{hyperref}
\usepackage{graphicx}
\usepackage[normalem]{ulem}
\usepackage{listings}
\usepackage{algorithm}
\usepackage[noend]{algpseudocode}
\usepackage{colortbl}
\usepackage{siunitx}
\usepackage{comment}
\usepackage{booktabs}
\usepackage[T1]{fontenc}        
\usepackage[utf8]{inputenc}     
\usepackage[table]{xcolor}
\usepackage{textcomp}

\usepackage{makecell}

\def\BibTeX{{\rm B\kern-.05em{\sc i\kern-.025em b}\kern-.08em
    T\kern-.1667em\lower.7ex\hbox{E}\kern-.125emX}}
\begin{document}

\newcommand{\hlnew}[1]{\textcolor{blue}{#1}}%

\title{Understanding the Power of Evolutionary Computation for GPU Code Optimization} 


\makeatletter
\newcommand{\linebreakand}{%
  \end{@IEEEauthorhalign}
  \hfill\mbox{}\par
  \mbox{}\hfill\begin{@IEEEauthorhalign}
}
\makeatother

\author{\IEEEauthorblockN{Jhe-Yu Liou}
\IEEEauthorblockA{\textit{Arizona State University} \\
Tempe, AZ \\
jhe-yu.liou@asu.edu}
\and
\IEEEauthorblockN{Muaaz Awan}
\IEEEauthorblockA{\textit{Lawrence Berkeley National Laboratory} \\
Berkeley, CA \\
mgawan@lbl.gov}
\and
\IEEEauthorblockN{Steven Hofmeyr}
\IEEEauthorblockA{\textit{Lawrence Berkeley National Laboratory} \\
Berkeley, CA \\
shofmeyr@lbl.gov}
\linebreakand %
\IEEEauthorblockN{Stephanie Forrest}
\IEEEauthorblockA{\textit{Arizona State University} \\
Tempe, AZ \\
stephanie.forrest@asu.edu}
\and
\IEEEauthorblockN{Carole-Jean Wu}
\IEEEauthorblockA{\textit{Arizona State University} \\
Tempe, AZ \\
carole-jean.wu@asu.edu}
}

\maketitle

\thispagestyle{plain}

\begin{abstract}


Achieving high performance for GPU codes requires developers to have significant knowledge in parallel programming and GPU architectures, and in-depth understanding of the application. This combination makes it challenging to find performance optimizations for GPU-based applications, especially in scientific computing.
This paper shows that significant speedups can be achieved on two quite different scientific 
workloads using the tool, GEVO, to improve performance over human-optimized 
GPU code.  
GEVO uses evolutionary computation to find code edits that improve the runtime of
a multiple sequence alignment kernel and a SARS-CoV-2 simulation by 28.9\% and 29\% respectively. Further, when GEVO begins with an early, unoptimized version of the sequence alignment program, it finds an impressive 30 times speedup---a performance improvement similar to that of the hand-tuned version. 
This work presents an in-depth analysis of the discovered optimizations, revealing that the primary sources of improvement vary across application; that most of the optimizations generalize across GPU architectures; and that several of the most important optimizations involve significant code interdependencies.
The results showcase the potential of automated program optimization tools to help reduce the optimization burden for scientific computing developers and enhance performance portability for domain-specific accelerators. 

\end{abstract}


\section{Introduction}
\label{sec:introduction}

Graphics Processing Units (GPUs) are widely used to accelerate parallel
applications in domains such as statistical modeling, machine learning, molecular simulations and bioinformatics, just to name a few. Although the tooling and programming language support for GPUs have matured, GPU programs remain challenging to optimize. 
Most GPU programs consist of parallel tasks, which compilers can optimize only to a certain extent, and issues such as thread mapping, communication, and synchronization are typically left for programmers to exploit manually. Consequently, applications often require hand-tuning to take full advantage of the GPU's computational power, which is time-consuming and requires significant expertise about parallel programming, underlying GPU architectures, and domain knowledge about the applications of interest.  

To tackle the aforementioned challenges,
prior works, such as~\cite{schkufza2013stochastic, chen2018tvm}, have explored automated compilation optimization methods to reduce the programming and performance optimization burden on application programmers.
One such approach uses evolutionary computation (EC) to optimize 
GPU programs represented in the LLVM~\cite{lattner2004llvm} intermediate representation (LLVM-IR)~\cite{GEVO}.  
An earlier study demonstrated that an EC-based approach
achieved run-time improvements on a wide variety of general-purpose, but mostly 
unoptimized GPU programs by an average of 51\%, performing especially well for error-tolerant applications. 
Despite these results, questions remain about \textit{what optimizations such a method can find}, \textit{how well it performs on hand-tuned production applications}, \textit{how the optimizations are discovered}, and \textit{how the method can be integrated into a production-level GPU application development}.

In this paper, we address these research questions by deploying GEVO and 
analyzing the performance optimization opportunities on two important bioinformatics applications: gene sequence alignment and a SARS-CoV-2 infection simulation. Aligning sequences of DNA, RNA or proteins is a fundamental operation in computational biology and underpins the success of many bioinformatics and medical applications~\cite{pareek2011sequencing}. The SARS-CoV-2 model (called \textit{SIMCoV}) simulates how virus interacts with the patient's immune system while spreading in a human lung and causing tissue damage. Accelerating the performance of the SARS-CoV-2 simulation is crucial for understanding the many complexities of COVID-19.

Both sequence alignment and the SIMCoV simulations are highly computation-intensive. 
For example, in the first six months of 2021, over 6.7 million CPU hours were used for genome assembly on National Energy Research Scientific Computing Cluster (NERSC)’s Cori Supercomputer, with roughly 40\% of the time spent in the sequence alignment kernel. Because of its importance, significant effort has been spent developing and manually optimizing ADEPT~\cite{awan2020adept}, a state-of-the-art GPU accelerated sequence alignment library. Similarly, it would take more than two weeks for SIMCoV to fully simulate a single infection, even for a single two-dimensional slice of human lung tissue, on a modern, consumer-level CPU.


We deploy GEVO on two versions of ADEPT, downloaded from its public open-source code repository. ADEPT-V0 is the version of the code before handtuning, whereas ADEPT-V1 represents the hand-optimized version. We show that the performance of ADEPT-V0 can be improved by 30 times on state-of-the-art GPUs---a level of performance that is similar to the hand-tuned version. On the hand-tuned  version (ADEPT-V1), an additional 28.9\% speedup is achieved with GEVO-discovered optimizations.  Similarly,  
GEVO finds optimizations providing 29\% performance improvement for the SIMCoV simulation code running on the P100 GPU.

We demonstrate that the benefits of automated program optimization tools are multi-dimensional, by using a tailored instrumentation of the program source code to localize the discovered optimizations and through a detailed performance analysis. 
Our results showcase the potential of automated program optimization tools to reduce the optimization burden for application developers, allowing them to focus on algorithms rather than details of hardware features and architecture specifics which are often black box or proprietary, and we show how such tools can actively influence the development of GPU application codes.

An important contribution of this work is its in-depth analysis of the discovered performance improvements, which can shed light on under-studied phenomenon. Our analysis shows that a key source of the impressive performance improvements are multiple interdependent code modifications, known as \textit{epistasis} in evolutionary biology.
To gain insight into how the search process assembles these interdependent code modifications, we recapitulate and analyze the search history from an informative run. 
We also convert the discovered code LLVM-IR modifications back to source code to characterize their contributions. To our knowledge, this is the first such study to reveal the importance of interdependencies in GPU code, which has implications for automated compiler optimization in general.

The contributions of the paper are summarized as follows:
\begin{itemize}
    \item While EC methods have been shown in prior work~\cite{GEVO} to improve performance of naive GPU programs, we demonstrate that these methods can compete directly with human experts, outperforming even hand-tuned GPU programs (Section~\ref{sec:result}).
    \item We conduct a detailed study and code analysis to characterize performance improvements and to explain how the optimizations were discovered and achieved. Compared to earlier EC-based work on software, which typically uses one or two mutations to repair small bugs or otherwise improve software, we find optimizations that involve many more mutations. 
    In some cases, a single GEVO optimized program contains nearly 1400 separate mutations (edits), of which up to 17 contribute significantly to the optimization. We define 
    a multi-step process to identify relevant interdependent clusters and report how they were discovered (Section~\ref{sec:edit_characteristic}). 
    \item We demonstrate the benefits of using EC methods in different program development stages, identifying performance hot-spots and strengthening a programmer's understanding of system performance improvement opportunities. The lessons learned can suggest further algorithmic improvements or manual adjustment of suggested optimizations, removing unwanted side effect if any.
    
\end{itemize}

By focusing on two computation-intensive workloads, 
our analysis reveals the importance of manipulating interdependencies to find performance enhancements at the LLVM-IR level, highlighting why stochastic methods like EC are particularly suitable 
for accelerating execution time performance of domain-specific computations beyond what is achievable by algorithm and hardware domain experts.

\section{Background}

This section briefly describes GEVO, provides relevant background on ADEPT and SIMCoV, and gives details about their corresponding GPU-based implementations.

\subsection{Evolutionary Search for GPU Code Optimizations}
\label{sec:gevo}

\begin{figure}
    \centering
    \includegraphics[width=0.9\linewidth]{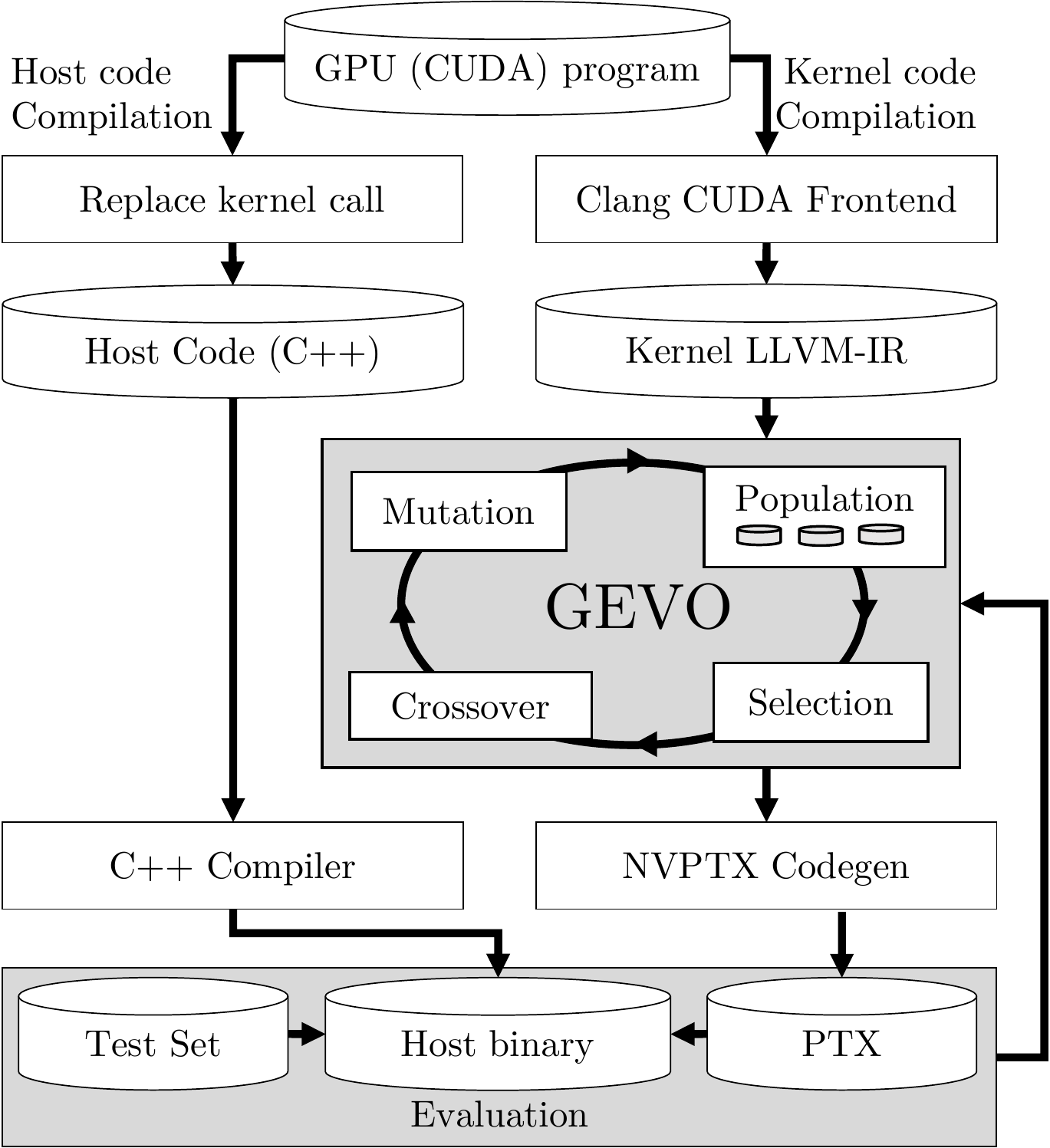}
    \vspace{-0.2cm}
    \caption{The GPU program compilation flow with GEVO interposed to dynamically modify and evaluate variants of the kernel code (gray blocks).} 
    \label{fig:cuda-compileflow}
\end{figure}

There is considerable interest in methods that automatically tune code after traditional compiler passes. 
Our work uses EC because it generalizes to large code sizes and can be applied generically to many software problems, including automated bug repair~\cite{genprog, yuanARJAAutomatedRepair2020}, energy reduction~\cite{gp4energy, bruce15energy}, and run-time optimization~\cite{GP4ProgImprovement, GP4cudaGzip}.
Many tools have been developed over the past decade for evolving program text~\cite{koza1994genetic, walsh1996paragen,genprog,gp4shader,marginean2019sapfix,yuanARJAAutomatedRepair2020}, and the vast majority of them operate on source code.  In a nutshell, these methods start with a single program, generate an initial population of program variants using random mutation operators, validate each variant by running it on multiple test cases, evaluate the valid variants according to a fitness metric, and use this information to select the best individuals, which are then subjected to further mutation and recombined with one another to produce novel variants.  This process is iterated until a time-out is reached or an acceptable solution is discovered.  Mutation operators are readily implemented in source code or assembly, but the single static assignment discipline of LLVM-IR complicates their implementation considerably.  The only mature EC tool that operates on LLVM-IR is GEVO (Gpu EVOlution)~\cite{GEVO}, which we adapted for the present work.  

GEVO takes as input a GPU program, user-defined test cases, and a fitness function to be optimized, which in our case is runtime.
Kernels that run on the GPU are first separated and compiled into LLVM-IR by the Clang compiler. GEVO takes these kernels 
as input, applies mutation and crossover to produce new kernel variants, and translates the implementations into PTX files. 
The mutations can either operate on an instruction (copy, delete, move, replace, or swap) or replace the operands between instructions. The host code running on the CPU is modified to load the generated PTX file into the GPU.  GEVO then evaluates the kernel variant according to the fitness function. This process is illustrated in Figure~\ref{fig:cuda-compileflow}. 

\subsection{Sequence Alignment}
\label{sec:ADEPT}
ADEPT implements Smith-Waterman, a widely used sequence alignment algorithm based on dynamic programming which guarantees an optimal \textit{local} alignment between two given sequences \cite{smith1981identification}.

\begin{figure}
    \centering
    \includegraphics[width=1\linewidth]{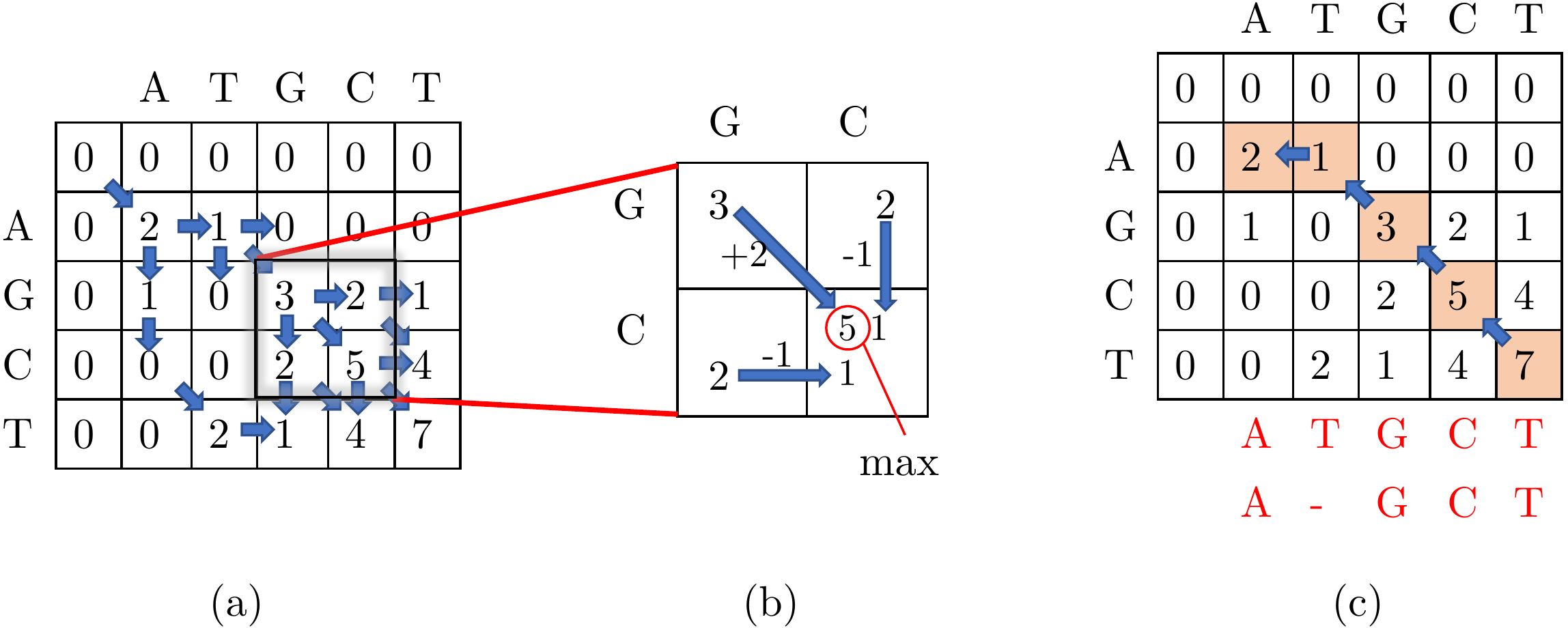}
    \vspace{-0.5cm}
    \caption{Example of the Smith-Waterman algorithm aligning two sequences, ATGCT and AGCT. (a) The forward pass calculates the scoring matrix with arrows showing how the scores are derived. (b) A single score calculation from the three neighboring cells. (c) The reverse pass from the calculated scoring matrix determines the alignment, with the final alignment result shown in the red text under the matrix.}
    \label{fig:smith-waterman}
\end{figure}

\subsubsection{Smith-Waterman Algorithm}
\sloppy Given two sequences $A=(a_1,a_2,...,a_n)$, $B=(b_1,b_2,...,b_m)$ to be aligned, 
a scoring matrix $H$ is calculated with size $(n+1) \times (m+1)$, where $n$ and $m$ are the length of $A$ and $B$ (Figure~\ref{fig:smith-waterman}(a)). The cell \(H_{ij}\) in the scoring matrix $H$ represents the highest alignment score with sequences ending in the pair of $a_i$ and $b_j$.

The cell score $H_{ij}$ is calculated by maximizing over the values from three directions of prior alignments ($H_{i-1,j-1}, H_{i,j-1}, H_{i-1,j}$) (Figure~\ref{fig:smith-waterman}(b)). The diagonal direction considers the similarity score $s$ of the current pair $a_i,b_j$ in the sequences, awarding the cell score (+2) if the paired $a_i,b_j$ is matched and
penalizing it (-2) otherwise. The vertical or horizontal direction introduces a gap in the current location of one
sequence or another. Gap insertion penalizes the cell score with a smaller penalty (-1) than a sequence pair mismatch. How the score is awarded or penalized is arbitrarily determined and can be changed based on particular scenarios.

After the scoring matrix is obtained by iterating the cell score calculation from top left to bottom right, the optimal alignment is generated by tracing back from the highest score in the matrix $H$, traversing along the highest score in the region in the reverse direction from how the matrix was calculated, until score zero is reached (Figure~\ref{fig:smith-waterman}(c)).

\subsubsection{GPU-accelerated Smith-Waterman Algorithm}
ADEPT parallelizes Smith-Waterman by offloading the computation of each column of the scoring matrix into one thread. As Figure~\ref{fig:adept-parallel} shows, the computation in each cell also depends on the scores of neighboring cells. Thus, the threads must be delayed, following the order of column index so the dependant values are ready to be shared from other threads. 

In the GPU CUDA programming model, developers can exchange thread data through global/host memory, GPU device memory, shared memory, or private-thread register~\cite{registercache}. The first two memory types have no restriction on which threads can exchange data, but data stored in the shared memory and private thread register are visible only within a thread block and a warp, respectively. 
Despite much faster data access latency, private registers are unfriendly to program because they involve low-level, intrinsic instructions. 
To reduce data movement latency, ADEPT optimizations exploit both shared memory and private registers for data exchange.

\begin{figure}
    \includegraphics[width=0.9\linewidth]{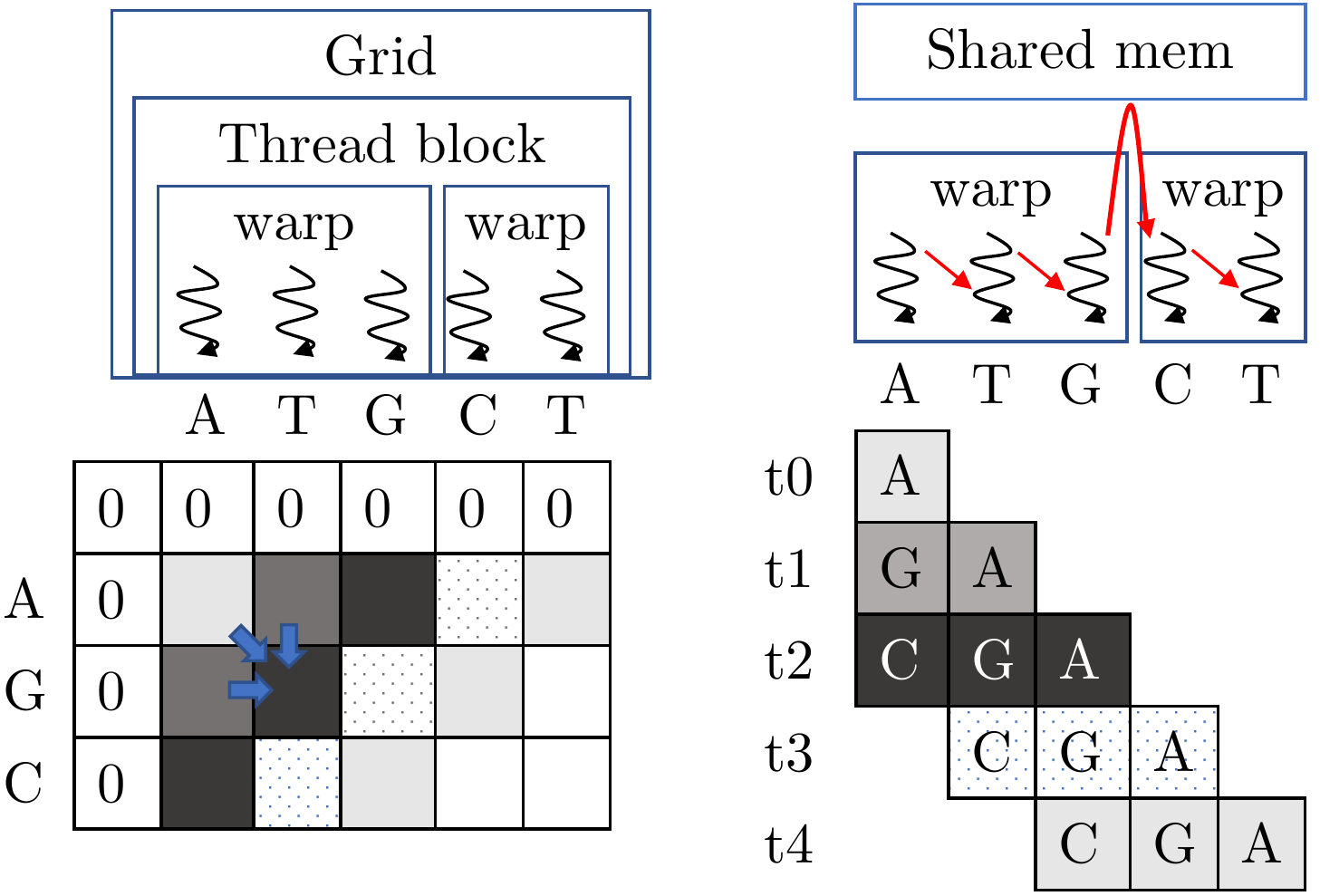}
    \vspace{-0.3cm}
    \caption{Illustration of the GPU-accelerated Smith-Waterman algorithm. 
    The kernel runtime performance can be improved depending on the data communication patterns: spatial (bottom, left) vs. temporal (bottom, right).}
    \label{fig:adept-parallel}
\end{figure}

\subsection{Coronavirus Simulation Model}
\label{sec:simcov}

Moses et al. developed a computationally intensive, spatially explicit model (SIMCoV) to study why infection trajectories vary so widely across different patients, even those with identical co-morbidities.~\cite{moses2021spatially}.  SIMCoV simulates both the spread of  virus (SARS-CoV-2) through the complex physical strucuture of the lung and important aspects of the immune response, modeling the dynamics of four important elements: epithelial cells, virions, inflammatory signals, and T cells. Given a simulation space, e.g., for simplicity, consider a grid that represents a two-dimensional slice of lung tissue, the model is initialized with an epithelial cell at each grid point, and a set of infection sites.
On each iteration, the model simulates four tasks for each grid point:
\begin{itemize}
    \item Circulating T cells \textit{extravasate} from the vascular system into the epithelial tissue with a probability determined by the presence of inflammatory signal.
    \item If the grid point contains a T cell, the T cell moves randomly to an adjacent location.
    \item Each epithelial cell’s state is updated to one of: healthy, infected, apoptotic (in the process of dying), dead. Virions cause healthy cells to become infected, and infected cells eventually die. T Cells trigger cell death by binding to cells, 
    preventing the further production of the virus.
    \item Virus and inflammatory signals diffuse from established sites of infection to neighboring grid points.
\end{itemize}

\subsubsection{GPU-accelerated SIMCoV}
SIMCoV's GPU implementation parallelizes its multi-core CPU implementation
to construct GPU kernels by assigning each grid point's calculation to a thread. Over 90\% of the GPU kernel runtime is spent moving T cells and spreading virus and inflammatory signals. 

\subsubsection{Stochastic Nature of the SIMCoV Simulation}
\label{sec:simcov:stochastic}
Many components of the model simulation are stochastic, e.g., T cell generation and movement. This mimics biology but also poses validation challenges for GEVO, which must determine the correctness of any code modification. Fixing the random seed removes most of the stochasticity, but not all. For example, the simulation does not allow two T cells to move into the same grid point, which can cause a race condition.
When such race conditions occur, the outcome is determined by the implementation of the GPU thread scheduler. This is an architecture-dependent approach and not transparent to application developers. 


\section{Experimental Setup}
\label{sec:Methodology}

\subsection{Compilation Preprocessing} 

First, we compile both the ADEPT and SIMCoV GPU kernels from CUDA into LLVM-IR using the Clang compiler. 
To enable code correspondence between the CUDA source and the GEVO-transformed codes, we instrumented the Clang compiler to enable source code debugging information and modified GEVO's mutation operator to encode the source code location information.  
Next, we modified both ADEPT's and SIMCoV's host code to invoke the GPU kernel from an external PTX file---the final product of the newly-mutated LLVM-IR that is executable by the CUDA binary.
The host code is compiled using NVIDIA's nvcc compiler~\cite{nvcc}.
Figure~\ref{fig:cuda-compileflow} illustrates the compilation process. 

\subsection{Application Code}

To study GEVOs effectiveness at  
different code development stages, we considered two versions of ADEPT:
\begin{itemize}
    \item {\bf ADEPT-V0} is the original parallel implementation  (423 lines of code from one CUDA kernel, 1097 LLVM-IR instructions) 
    \item {\bf ADEPT-V1} is a manually-optimized version by an expert in both the application domain and GPU
    (623 lines of code from two CUDA kernels, 1707 LLVM-IR instructions). 
\end{itemize}

ADEPT-V1 contains NVIDIA hardware-specific intrinsics, which
use both shared memory and private registers for data exchanges (Section~\ref{sec:ADEPT}).  
ADEPT-V1 executes approximately 20-30 times faster than ADEPT-V0 across the GPUs used in this paper.  

For SIMCoV, the only available GPU code to us was an initial GPU port from its multi-core CPU implementation, similar to ADEPT-V0, with 1197 line of code from 8 GPU kernels, translating to 1712 LLVM-IR instructions.

\subsection{ Validating Code Transformations} 

For ADEPT, We used the 30,000 pairs of DNA gene sequences in the ADEPT repository
for fitness evaluation.
In addition, we held out 4.6 million pairs of sequences to validate the final optimized ADEPT code. 
Although GEVO can trade off error tolerance against performance objectives, gene sequence alignment often requires strict accuracy so we require 100\% accuracy for our ADEPT validation.

SIMCoV does not have a formal testing dataset for verification. Therefore, we controlled the simulation environment by fixing the initial random seed so that the simulation progress, including virus spread, epithelial cell state, and number of T cells is as  similar as possible across runs. We use the simulation output generated from the unmodified SIMCoV as ground truth. 
To manage the remaining non-determinism, we introduce the concepts of per-value mean and per-value variance to measure how close the output is to ground truth.

To evaluate fitness of a SIMCoV variant, we run the simulation on a small, 100x100 grid for 2500 simulation steps, which is generally insufficient for the simulation to reach steady state. Similar to ADEPT's held-out tests, we further validate the final GEVO optimized SIMCoV program after the run by both running the same 100x100 grid size for 10,000 simulation steps and by simulating a much larger, 2500x2500, grid. We were unable to run our optimized SIMCoV on a 10,000x10,000 grid, as the original paper did, due to size limits of the GPU memory.


\begin{table}
\centering
\caption{Architectural characteristics of the GPUs}
\vspace{-0.3cm}
\label{table:gpus}
\begin{tabular}{ llll}
\toprule
\textbf{GPU}  & \textbf{P100} & \textbf{1080Ti} & \textbf{V100} \\
\midrule
\rowcolor[HTML]{EFEFEF}
\begin{tabular}[c]{@{}l@{}}Architecture \\ Family\end{tabular} & Pascal & Pascal & Volta \\
 CUDA cores & 3584 & 3584 & 5120 \\
\rowcolor[HTML]{EFEFEF}
\begin{tabular}[c]{@{}l@{}}Core Frequency \end{tabular} & 1386 Mhz & 1999 Mhz & 1530 Mhz \\
\begin{tabular}[c]{@{}l@{}}Memory Size \end{tabular} & 16GB HBM & 11GB GDDR5X & 16GB HBM2 \\
\bottomrule
\end{tabular}
\end{table}

\subsection{System Hardware}

We evaluated and analyzed performance improvement using three generations of NVIDIA GPUs: P100~\cite{p100gpu}, 1080Ti GPU~\cite{1080tigpu}, and V100~\cite{v100gpu}, 
summarized in Table~\ref{table:gpus}. We disabled the GPU Boost Technology~\cite{gpuboost} to maintain constant GPU operating frequency for the experiments. The machine with P100 GPU has a 20-core CPU with 256GB memory. For V100 GPU, we used NERSC's Cori Supercomputer's GPU partition \cite{gpu_nodes}. In most cases, these runs used one V100 GPU with 10 CPU cores and 16GB memory.

\subsection{GEVO Specification} 

Kernel execution time is the fitness target, averaged across all test cases.
Individuals that fail one or more test cases are not part of the calculation.
We set the population size to 256, retained the four best individuals into the next generation (elitism), applied crossover with 80\% probability for each individual, and used a mutation probability of 30\% per individual per generation. Different search budgets are given to GEVO for ADEPT (7 days) and SIMCoV  (2 days), which roughly translates into 300 and 130 generations respectively. 

\section{Performance Evaluation Results}
\label{sec:result}

\begin{figure}
    \centering
    \includegraphics[width=1\linewidth]{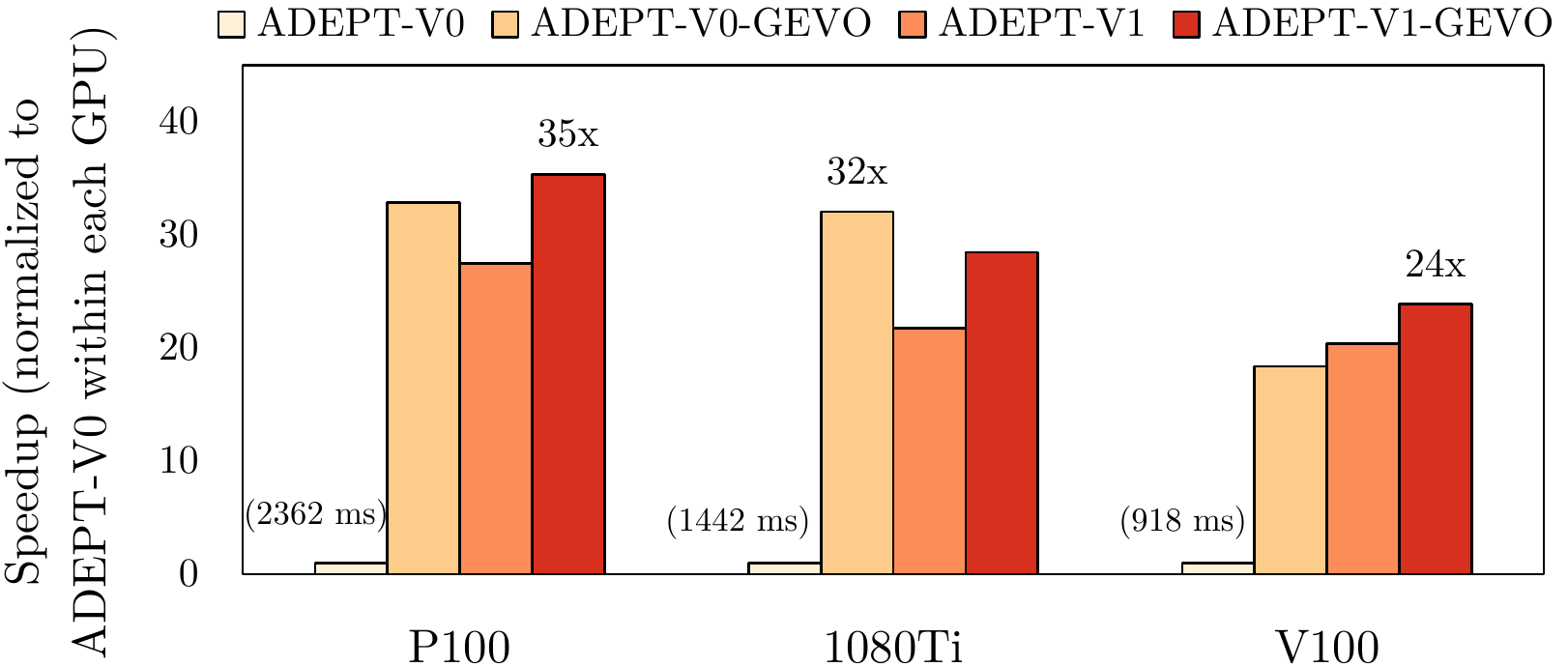}
    \vspace{-0.6cm}
    \caption{The performance results of ADEPT on the three generations of the GPUs.}
    \label{fig:result_adept}
\end{figure}

\noindent {\bf Summary:} 
Figures~\ref{fig:result_adept} and~\ref{fig:result_simcov} present the performance improvements for ADEPT-V0, ADEPT-V1, and SIMCoV on three generations of the GPUs.
Execution time improved for ADEPT-V0 by 32.8X, 32X, and 18.36X on the P100, 1080ti, and V100 GPUs, reducing the kernel runtime from 2,362 ms to 72 ms, from 1442 ms to 45 ms, and from 918 ms to 50 ms, respectively. For the hand-tuned, well-optimized version, ADEPT-V1, 
GEVO finds an optimization that achieves 1.28X, 1.31X, and 1.17X performance improvement on the P100, 1080ti, and V100 GPUs. 
For SIMCoV, the performance improvement is 1.29X, 1.42X, and 1.16X on the P100, 1080ti, and V100 GPUs, respectively.

Because GEVO implements a stochastic search, we next ask how much variation there is across the experimental runs. 
Because the experiments are computationally expensive, we focused our analysis on 
on the P100 GPU and conducted ten independent runs for each configuration (Figure~\ref{fig:ten_runs}). 
For ADEPT-V1, compared with the initial run (1.29X improvement indicated by the solid blue line in Figure~\ref{fig:ten_runs}(a)), the highest speedup is 1.33X while the lowest is 1.1X. The mean is 1.20X and the variance is $\pm0.08$. Figure~\ref{fig:ten_runs}(b) shows that, for SIMCoV, the highest speedup is 1.35X and the lowest is 1.18X, with a mean of 1.28X and variance of $\pm0.06$.
These results convey the value of running GEVO multiple times to discover the best possible optimization.
The sources of the performance improvement for the ADEPT and SIMCov GPU codes are distinct, which we analyze and present in detail in Section~\ref{sec:edit_alyz}.

\begin{figure}
    \centering
    \includegraphics[width=1\linewidth]{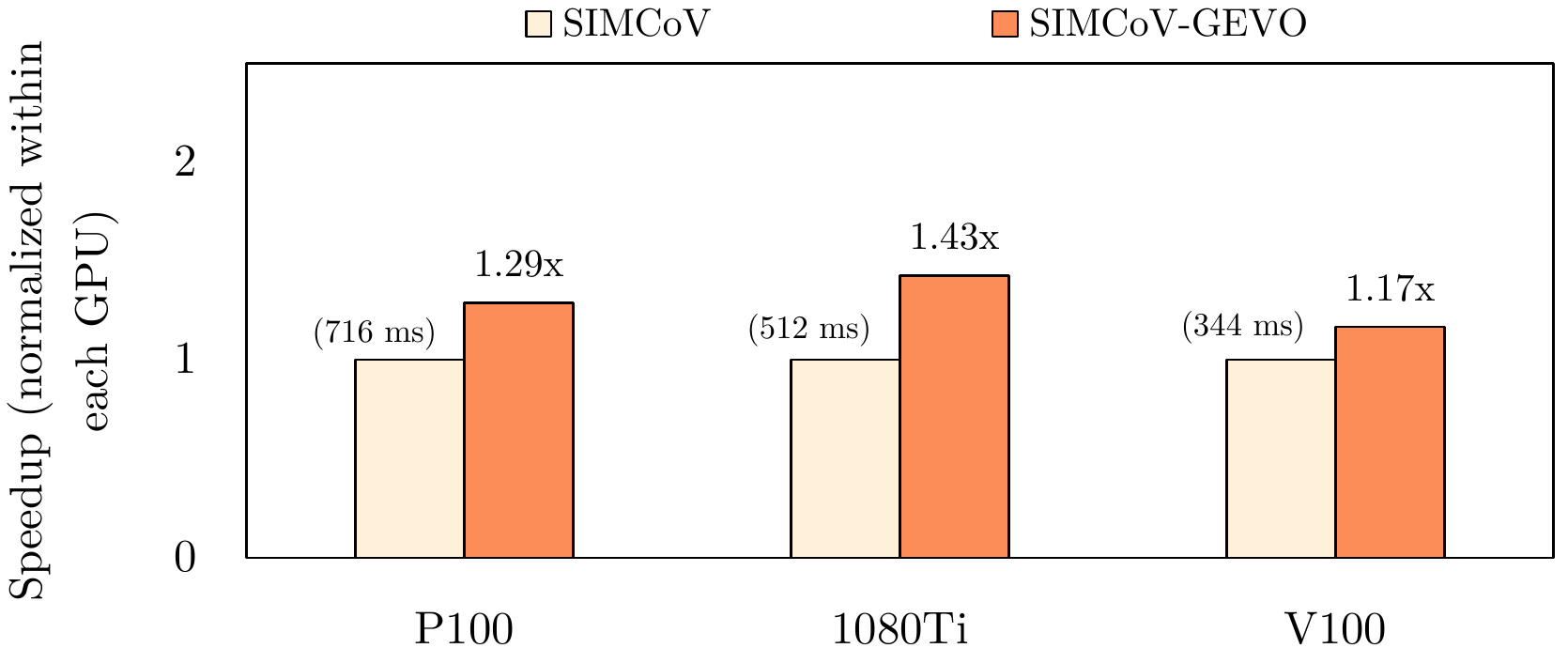}
    \vspace{-0.5cm}
    \caption{The performance results of SIMCoV on the three generations of the GPUs.}
    \label{fig:result_simcov}
\end{figure}

\begin{figure*}
    \centering
    \includegraphics[width=1\textwidth]{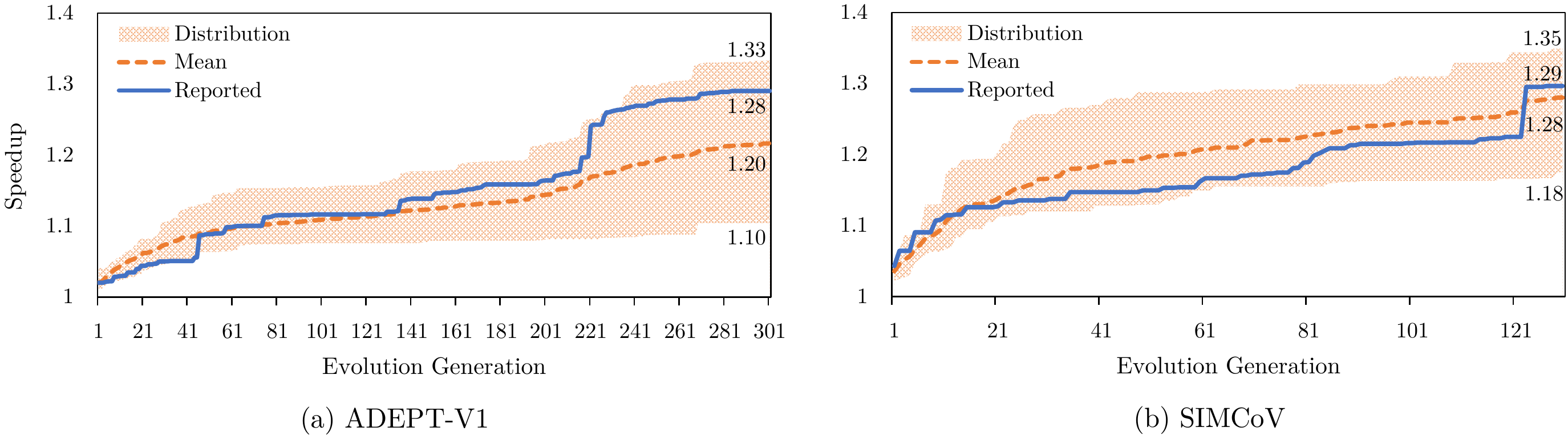}
    \vspace{-0.6cm}
    \caption{Distribution of performance improvements across ten GEVO runs for (a) ADEPT-V1 and (b) SIMCoV on P100. The shaded area encloses the historical path for all runs, while the dashed line indicates the average. 
    }
    \label{fig:ten_runs}
\end{figure*}

\noindent {\bf Generality:} 
To assess portability of the discovered optimizations, we ran ADEPT-V0) (GEVO optimized for the P100) on the V100 GPU and compared its performance to ADEPT-V) GEVO optimized for the V100.
The former achieves 99\% of the performance gain of the latter and similarly for the other 1080 Ti GPUs, suggesting that many of the optimizations generalize across the three GPUs which feature distinct compute and memory architectures. 
We observed similar performance portability with SIMCoV.
However, with ADEPT-V1, the same analysis showed that a small subset of the optimized code from the P100 GPU cannot run directly on the V100 GPU, suggesting that some performance optimizations are GPU architecture-dependent. 

\section{Understanding the Optimizations}
\label{sec:edit_characteristic}
To study the GEVO-discovered optimizations,
we define a multi-step process,
which first eliminates edits that contribute less than 1\% performance
improvement (\textit{weak} mutations), then separate out mutations (edits) that are
independent, i.e. those that achieve greater than 1\% fitness improvement
independent of the other edits in the set.  We can then conclude that
the remaining mutations are interdependent (epistatic), but we do not
know if the entire set is mutually interdependent, or if
there are subsets. 
To find the subsets, we conduct an exhaustive search of
all possible combinations of the epistatic edits, which is feasible because the total number of epistatic edits is small.
(For example, the edit number is reduced to 12 from 1394 on ADEPT-V1)
The following subsections describe each step in detail, primarily using ADEPT-V1 and SIMCoV on P100 as examples. 

\subsection{Edit Minimization}
\label{sec:weak_edits}
Overall, the best performing code variants from ADEPT-V1 and SIMCoV on a P100 GPU contained a total of 1394 and 384 mutations, respectively.
It is surprising that
the code is robust against so many mutations while preserving required functionality. To focus on the performance-critical changes, and to avoid side effects, we removed weak edits from consideration (Algorithm~\ref{alg:weaks}). 

\algnewcommand\algorithmicforeach{\textbf{for each}}
\algdef{S}[FOR]{ForEach}[1]{\algorithmicforeach\ #1\ \algorithmicdo}

\def\BState{\State\hskip-\ALG@thistlm}
\makeatother
\begin{algorithm}[]
\caption{Identify weak edits }\label{alg:weaks}
\begin{algorithmic}[1]
\Statex \textbf{Parameter}: Edit set $S=\{e_1,...,e_n\}$
\Statex \textbf{Function} $f(S)$: measure the fitness (performance) of the program with edit set $S$ applied
\State $weaks \gets \emptyset$
\ForEach{$e_i \in S$}
  \If{$\dfrac{f(S-weaks) - f(S-weaks-e_i)}{f(S-weaks-e_i)} < 1\%$}
    \State $weaks \gets weaks + e_i$
  \EndIf
\EndFor
\end{algorithmic}
\vspace{-0.1cm}
\end{algorithm}

We systematically measured the performance difference between the program variant with and without each target mutation, in the context of all the remaining mutations.
Any individual edit may not have immediate impact on kernel execution time, but it could enable other higher-performing program mutants, serving as a kind of stepping stone. Our systematic reduction identifies these false-negative cases for weak edits.
It is also possible that  when weak edits are removed from consideration, 
we avoid a situation where multiple weak edits can potentially lead to the identical program variant. 
For example, suppose edits $e_1$ and $e_2$ are both stepping stones leading to $e_3$. In this case, $e_1$ and $e_2$ are redundant, and one of the two can be safely removed from the edit set without performance impact. 
We measure the 1\% performance threshold 
using the \textit{nvprof} profiling tool. This process reduces the number of code edits in our set from 1394 to 17 for ADEPT-V1 with minimal reduction of performance (0.9\%), corresponding to performance improvement of 28\% instead of 28.9\%.

\subsection{Edit Interactions}

\newcommand{\algorithmiccontinue}{\textbf{continue}}
\newcommand{\CONTINUE}{\STATE \algorithmiccontinue}

\def\BState{\State\hskip-\ALG@thistlm}
\makeatother
\begin{algorithm}
\caption{Separate independent and epistatic edits. }\label{alg:epistasis}
\begin{algorithmic}[1]
\Statex \textbf{Parameter}: Edit set $S=\{e_1,...,e_n\}$
\Statex \textbf{Function} $f(S)$: measure the fitness (performance) of the program with edit set $S$ applied
\State $Indep \gets \emptyset$
\ForEach{$e_i \in S$}
    \If{$f(e_i)$ or $f(S-Indep-e_i)$ fails}
        \State \textbf{continue}
    \EndIf

    \State $PerfIncr\gets \dfrac{f(\emptyset)-f(e_i)}{f(\emptyset)}$
    \State $PerfDecr\gets \dfrac{f(S-Indep-e_i)-f(S-Indep)}{f(S-Indep-e_i)}$
    \If{$PerfIncr \simeq PerfDecr$}
        \State $Indep \gets Indep + e_i$
    \EndIf
\EndFor
\State $Epistasis \gets S - Indep$ 
\end{algorithmic}
\vspace{-0.1cm}
\end{algorithm}
\vspace{-0.1cm}

Next, we describe how to 
identify interactions (epistasis) among edits, producing a set of independent edits and a set of epistatic edits 
(Algorithm~\ref{alg:epistasis}).  The  algorithm first identifies the set of independent edits, and whatever remains after the procedure is considered to be epistatic. An independent edit must individually be both applicable and removable from the edit set (lines 4 and 5 of Algorithm~\ref{alg:epistasis}) without causing an error. If it passes this check, we next evaluate how performance changes with and without the edit applied, first to the  empty set of edits (i.e. to the original program) and then in the context of the remaining edit set (lines 6 to 9 of Algorithm~\ref{alg:epistasis}). If the run-time from the above two tests agrees, the edit is identified as independent.  
In our running example, this algorithm divided the 17 significant edits from Section~\ref{sec:weak_edits} into 5 independent and 12 epistatic edits. The two sets contribute 7\% and 17\% performance improvement to ADEPT-V1, respectively. Interestingly, we did not find performance-impactful epistatic edits for ADEPT-V0 or SIMCoV.

\begin{figure}
    \centering
    \includegraphics[width=1\linewidth]{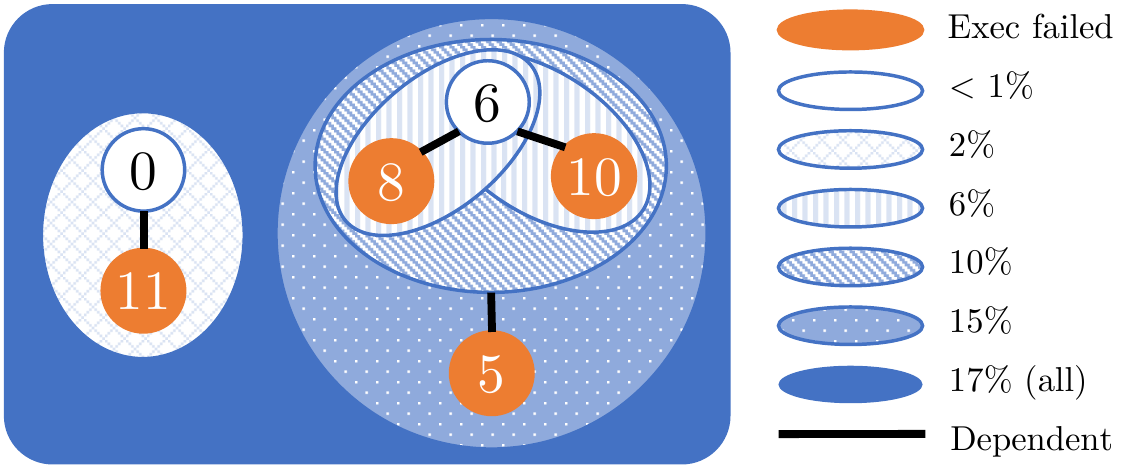}
    \vspace{-0.5cm}
    \caption{The edit relation graph and corresponding performance improvements for GEVO-optimized ADEPT-V1 on P100 GPU. Each node is an individual edit labeled with its index. The different backgrounds show  the performance improvement for the different edit combinations, where orange color indicates execution failure when certain edits when applied individually, like edit 8.} 
    \label{fig:epistasis_relation}
\end{figure}

\begin{figure}
    \centering
    \includegraphics[width=1\linewidth]{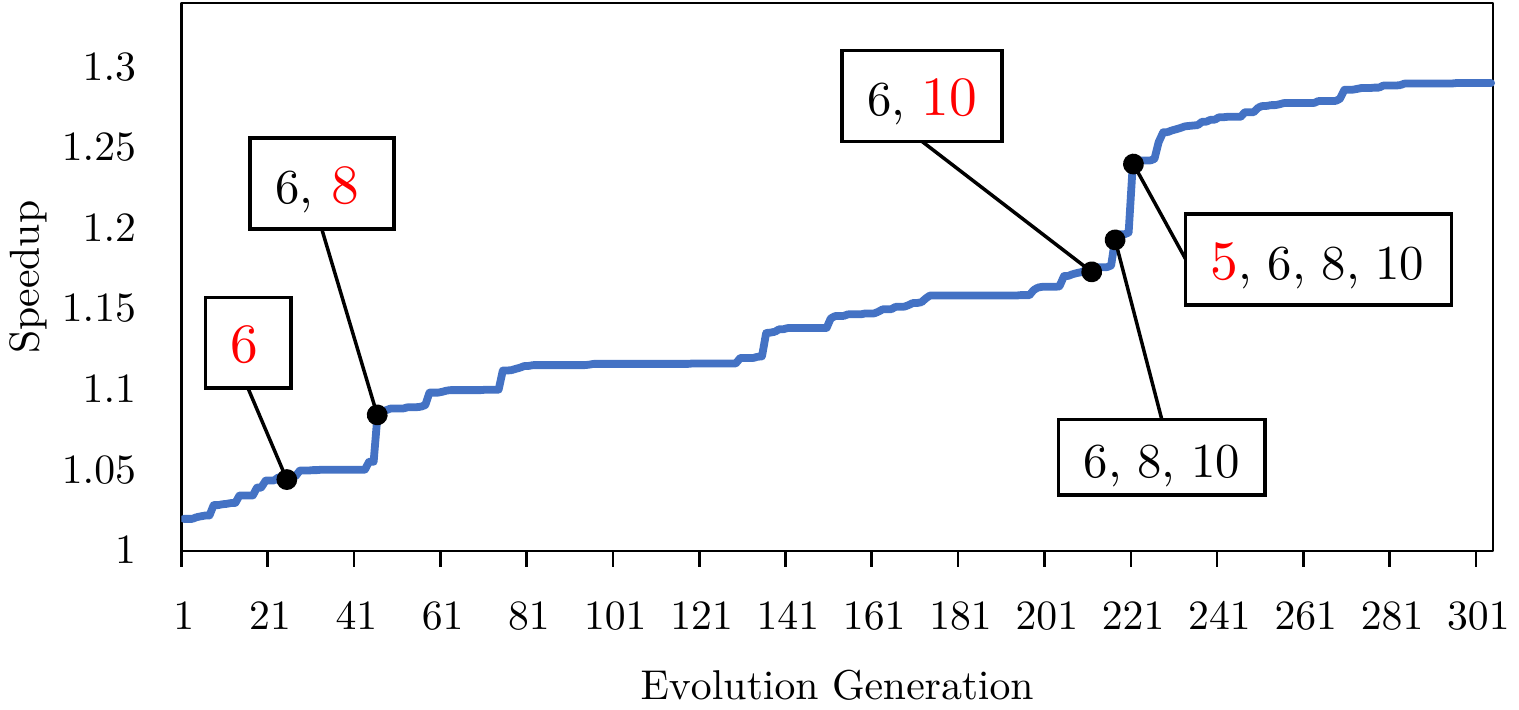}
    \vspace{-0.5cm}
    \caption{The discovery sequence for edits in the epistasis set (edits 5, 6, 8, and 10) across 303 generations. These are the same edits to ADEPT-V1 on P100 GPU shown in Figure~\ref{fig:epistasis_relation}. The group of edits in each box indicates in which generation this group was found, and  edits colored red indicate the first time that edit was discovered. }
    \label{fig:epistasis_generation}
\end{figure}

\label{sec:epistasis_alyz}
\subsection{Epistatic Edit Set Analysis}
While prior work in EC for software improvement rarely discovers epistasis (e.g., in bug repair it is usually one or two mutations), epistasis is common in biology~\cite{epistasis}. 
We analyze the epistatic set for ADEPT-V1. This set consists of twelve edits. We show a dependency graph (Figure~\ref{fig:epistasis_relation}) for the most important epistatic clusters---determined by evaluating every subset of the epistatic set.
The numbers in circles represent the edit index, and the black lines indicate a dependency relation. 


There are two independent epistatic subgroups. One subgroup (edits 5, 6, 8, and 10) is the most significant, contributing 88.2\% of the overall 17\% performance improvement. Edits 8 and 10 both depend on the success of edit 6. The program mutants with either edit 8 or edit 10 individually fail the verification step. Edit 5 also fails individually and requires all three remaining edits (6, 8, and 10), to function properly.  We consider this most significant cluster of edits in detail.
Figure~\ref{fig:epistasis_generation} shows shows when the edits were discovered and how the discovery affected fitness.
As expected, edit 6 with no dependencies was discovered first, followed by edit 8 in the 47th generation, edit 10 in the 213th generation, and edit 5 in the 221st generation.

The performance variation from run to run (figure~\ref{fig:ten_runs}(a)), was affected by the completeness of the discovered epistatic subgroups. For example, in the best run, GEVO further expanded the epistatic subgroup (e0, e11) to a 4-edit cluster similar to the subgroup (e5, e6, e8, e10). In the lowest performing run, GEVO discovered (e6, e10) but missed e8 and e5. 

\section{Functional Analysis of the Optimizations}
\label{sec:edit_alyz}
This section explores the functional impact 
of the key mutations from Section~\ref{sec:edit_characteristic}. We do so by tracing each relevant code edit in the LLVM-IR level back to its corresponding CUDA source code. 
Although requiring significant manual effort, this is an important step in understanding the performance optimization opportunities that EC can uncover. 

\definecolor{backcolour}{rgb}{0.95,0.95,0.92}
\definecolor{codegreen}{rgb}{0,0.6,0}
\definecolor{codegray}{rgb}{0.5,0.5,0.5}
\definecolor{codepurple}{rgb}{0.58,0,0.82}
\definecolor{backcolour}{rgb}{0.95,0.95,0.92}

\begin{figure}
\lstdefinestyle{mystyle}{
    language=C++, escapechar=|,
    morekeywords = {__shared__, __syncthreads, __shfl_sync},
    otherkeywords = {__shared__, __syncthreads, __shfl_sync},
    backgroundcolor=\color{backcolour},
    keywordstyle=\bfseries\color{blue},
    numberstyle=\ttfamily\small\color{codegray},
    commentstyle=\color{codegreen},
    basicstyle=\footnotesize\ttfamily,
    numbers=left,
    frame=single,framexleftmargin=3em,
    stepnumber=1,
    showstringspaces=false,
    tabsize=1,
    breaklines=true,
    breakatwhitespace=false,
    xleftmargin=8.0ex,
    moredelim=[is][\bfseries\color{red}]{<<<}{>>>},
}
\lstset{style=mystyle}
\begin{lstlisting}
...
// if (laneId == 31)
<<<if (landId == 0) { >>>// edit 5 |\label{line:e5}|
  sh_prev_E[warpId] = _prev_E;
  sh_prev_prev_H[warpId] = _prev_prev_H;}

// if(diag >= maxSize)
<<<if (tID < minSize) { >>>// edit 6 |\label{line:e6}|
  local_prev_E[tID] = _prev_E;
  local_prev_prev_H[tID] = _prev_prev_H; }

__syncthreads();

if (is_valid[tID] && tID < minSize) { |\label{line:exp1}|
  ...
  // if(diag >= maxSize) {
  <<<if (is_valid[tID]) >>> // edit 8 |\label{line:e8}|
    eVal = local_prev_E[tID-1] + extendGap;
  else { |\label{line:else1}|
    if (warpId != 0 && landId == 0)
      eVal = sh_prev_E[warpId-1];
    else // private register
      eVal = __shfl_sync(...); } |\label{line:end_e8}|

  // if(diag >= maxSize) {
  <<<if (is_valid[tID]) >>> // edit 10 |\label{line:e10}|
    final_H = local_prev_prev_H[tID-1];
  else { |\label{line:else2}|
    if (warpId != 0 && landId == 0)
      final_H = sh_prev_prev_H[warpId-1];
    else // private register
      final_H = __shfl_sync(...); |\label{line:end_e10}| 
  } ...

\end{lstlisting}
\vspace{-0.25cm}
\caption{Simplified code snippet from ADEPT-V1 for how data is exchanged using both private registers and shared memory. In edits 5, 6, 8, and 10 (red text, lines~\ref{line:e5}, \ref{line:e6}, \ref{line:e8}, and \ref{line:e10}), GEVO eliminates private registers and uses shared memory instead.}
\label{fig:data_exchange}
\end{figure}

\subsection{Rearrange Usage of Sub-Memory Systems on GPU}
\label{sec:result:submemory}
The epistatic edits identified in Section~\ref{sec:epistasis_alyz} alter how ADEPT-V1 uses the GPU's shared memory and private registers. By doing so, 15\% performance improvement is achieved on the P100. These edits are applicable on the V100 as well, achieving similar performance improvement. Recall that, in Section~\ref{sec:ADEPT}, ADEPT-V1 uses both private registers and shared memory to exchange data. Its implementation is shown in Figure~\ref{fig:data_exchange} with GEVO mutations indicated in red. These edits essentially eliminate the use of private registers and rely only on shared memory. 

The $else$ clauses at lines~\ref{line:else1} and \ref{line:else2} are for the thread that meets the conditions to share data through private registers using the $shfl\_sync$ function. Due to a limitation of the GPU architecture, GPU threads that cannot exchange data through private registers communicate through shared memory. The effect of edits 8 (line~\ref{line:e8}) and 10 (line~\ref{line:e10}) is to drop the use of private registers. It is achieved by replacing the corresponding $if$ condition with the existing boolean expression from line~\ref{line:exp1}. If the boolean expression in line~\ref{line:exp1} is true, both lines \ref{line:e8} and \ref{line:e10} are evaluated as true. This effectively causes every relevant GPU thread in the code snippet to write/read the data to/from the shared memory regardless of any other condition. However, edits 8 and 10 cannot be applied alone without edit 6 that implicitly enables every thread writing its data to the shared memory named \textit{local\_prev\_XX}.  After applying the three aforementioned edits, the shared memory named \textit{sh\_prev\_XX} is not required, leading to edit 5. At this stage, a human developer would likely remove the entire $if$ clause at lines~\ref{line:e5} since the shared memory within the if clause is no longer referred to. Instead of removing the shared memory, edit 5 is introduced that only changes which thread will access the shared memory. This modification achieves the same performance improvement as if the affected code snippet were removed. We suspect that by changing the memory access pattern, as edit 5 does, the GPU can schedule the memory access differently to hide the memory latency of this particular access~\cite{lee:ispass2014}.

Accessing private registers on GPUs is much faster than the shared memory. So then, how do edits that leverage shared memory achieve performance advantage? This might be related to branch divergence. Recall from Section~\ref{sec:ADEPT} and Figure~\ref{fig:adept-parallel}, while some threads in a warp can use private registers for data sharing, there is often one thread, usually the first thread in the warp, that must communicate through shared memory. Combining with the GPU lock-step execution model, i.e., every thread in the same warp executes the same instruction at the same time, the aforementioned behavior guarantees branch divergence in the if-else region between lines \ref{line:e8}-\ref{line:end_e8} and \ref{line:e10}-\ref{line:end_e10}. This essentially forces every thread in the same warp to run through both if and else regions, and whichever thread uses private registers has to wait for the slowest thread that accesses the shared memory to finish. As a result, the advantage of the fast access latency using the private registers is lost.

\begin{figure*}
    \centering
    \includegraphics[width=0.9\textwidth]{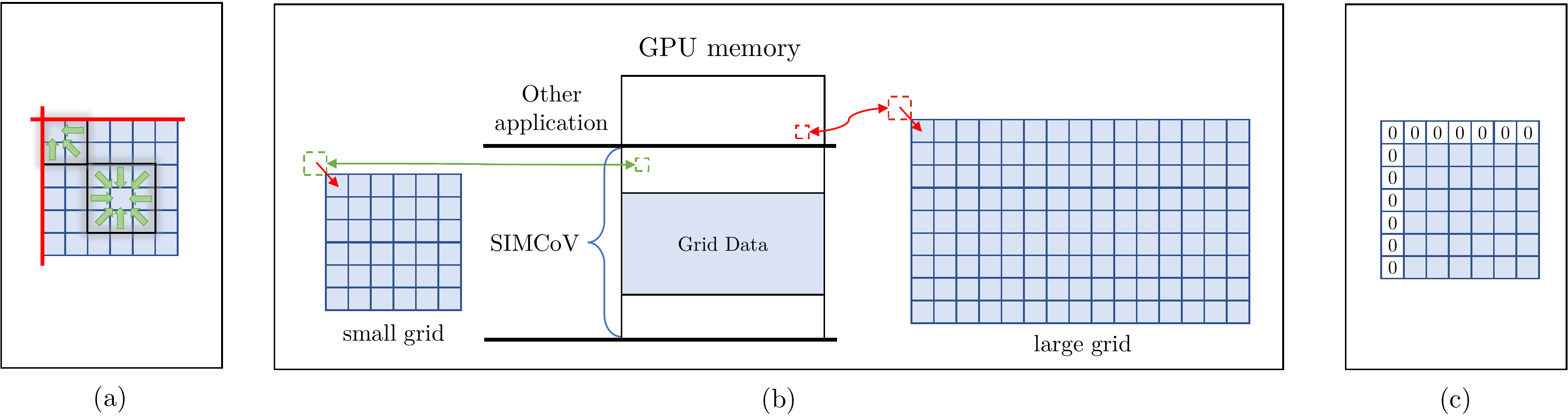}
    \vspace{-0.25cm}
    \caption{(a) illustrates that boundary check is a necessary step in the SIMCoV code. (b) illustrates how the boundary check removal is acceptable in a small grid but would fail for a large grid, which can be resolved by (c) padding the grid borders with extra grid points of 0 manually.}
    \label{fig:simcov_boundary}
\end{figure*}

\subsection{Remove Warp-Level Synchronization}
\label{sec:result:sync}
The CUDA programming guide suggests that, before exchanging data through the private register, programmers should invoke a query function, such as \textit{activemask} or \textit{ballot\_sync}, in order to return a mask indicating which threads are still alive in the warp. In particular, after the NVIDIA Volta GPU architecture (V100 GPU in our evaluation environment), \textit{ballot\_sync} should be used as the query function inside any conditional branch where branch divergence can happen. The reasoning is that the Volta architecture allows GPUs to subdivide a warp into subgroups to be scheduled independently, and \textit{ballot\_sync} implicitly forces the GPU to synchronize threads in the same warp. 

Perhaps to be conservative, the developers of ADEPT used both \textit{activemask} and \textit{ballot\_sync} before accessing the private registers in a conditional branch. An independent edit shows that removing \textit{ballot\_sync} yields 4\% performance improvement on the V100 GPU but not on the P100 GPU. This supports the idea that \textit{ballot\_sync} performs warp-level synchronization on the Volta GPU architecture but not on the older GPU architectures. 
This edit is interesting because it violates the CUDA programming guide~\cite{warplevelprimitives}.  Yet, 
the edit passes all the verification tests. However, due to the proprietary design of the Volta GPU warp scheduler, we cannot conclude in which situations it is safe to remove warp-level synchronization. 

\subsection{Remove Unnecessary Memory Initialization and Synchronization Procedures}
\label{sec:result:mem_init}
For ADEPT-V0, BEVO removed a small code region consisting of \textit{memset} and \textit{syncthred} functions for shared memory initialization and synchronization. This change improved the kernel performance by more than thirty-fold. In this case, it appears that we can completely ignore shared memory initialization,
even on the algorithm level, because other edits were not engaged to compensate for the behavior change. In fact, the human expert also removed this code region in ADEPT-V1.
In this case the initialization is required, but the way it is implemented is vastly inefficient. The original code asks all the GPU threads to perform memory initialization on the same memory region. Combined with synchronization,
GPU threads block each other to initialize the same memory region over and over again, creating the significant performance bottleneck. The common practice is to initialize the memory through the CUDA API outside the kernel or through the in-kernel code using only one active thread.
For application developers, the ability to quickly identify promising performance hot-spots that are challenging to discover using conventional tools is valuable, and this example highlights how GEVO supports this task.

\subsection{Boundary Check Removal and Grid Padding}
\label{sec:result:boudary}
In SIMCoV, GEVO removed multiple conditional branches, which disabled a grid boundary check. Its purpose is prevent erros when accumulating inflammatory signals from the neighboring grid points (the fourth task in Section~\ref{sec:simcov}). As Figure~\ref{fig:simcov_boundary}(a) shows, the boundary check  prevents the edge grid points from attempting to accumulate values from points outside of the grid (illegal memory accesses).
The performance analysis presented in this section addresses the following questions: \textbf{(1)} The boundary check optimization alone achieves 20\% performance improvement. How does a simple boundary removal achieve such disproportional execution time improvement? \textbf{(2)} How can out-of-bound memory access not break the program's behavior? 

To answer the first question, we examined the kernel with the modified code region. Surprisingly, a significant portion (31\%) of the kernel instructions were performing logic operations related to the boundary comparison, although, as shown in Figure~\ref{fig:simcov_boundary}(a), the vast majority of the grid points are not located on the boundary. Removing the boundary check, however, is only legitimate if there is a compensating code modification to prevent illegal accesses outside the boundary. This example demonstrates how the GEVO approach can inform application developers. By actively searching through the code for performance optimization opportunities, the search can expose promising performance hot-spot regions that may be overlooked otherwise.

We answer the second question using validation test sets. That is, by running the SIMCoV simulation at a larger grid size: 2500x2500. Even though the SIMCoV code passes the initial test using a smaller simulation area, the boundary check optimization triggers a segmentation fault on this larger held-out test (Figure~\ref{fig:simcov_boundary}(b)). 
It is not surprising that larger held-out tests are needed during the optimization search process
to detect such out-of-bound memory accesses, and this is a routine part of our evaluation strategy.
After probing the code and the boundary check optimization more deeply, we observed that simply padding the grid borders with extra points of value 0 (Figure~\ref{fig:simcov_boundary}(c)), the application can achieve a 14\% performance improvement with a negligible increase in the memory requirement.

\subsection{Remaining Edits}
\label{sec:remaining}
We attempted to analyze all the edits, but there are some that we were unable to decipher. For example, one edit duplicates a memory write operation to a region that no subsequent code ever accesses. Such an operation seems redundant and could slow down program runtime. Surprisingly, it improves the kernel performance by 1\% when run on the P100 GPU.
\section{Discussion}
Based on the mutational edit analysis described in Section~\ref{sec:edit_alyz}, many relevant mutations are related to the GPU architecture. This implies that, although the GPU programming model has matured in the past decade or two, it is still difficult to master, especially for 
hardware-related programming language features. Scientific applications, such as those we consider here, are often written by domain experts who are not necessarily trained as software developers. In these circumstances, an approach such as GEVO is an interesting and promising choice for GPU code optimization. We contacted the original developers of both ADEPT and SIMCoV, presented the discovered optimizations, and asked for their opinions and feedback. The developers were surprised that EC could synthesize code modifications with such large performance improvements. The main developer of ADEPT told us, "\textit{If I was aware such an automatic optimization tool existed, it might have saved a couple of months of effort, especially for optimizing toward a specific GPU architecture!}" And, from the developer of SIMCoV, "\textit{When I looked at the optimizations found for SIMCoV, I saw how I could change my algorithm to improve its
performance at scale. On CPUs, SIMCoV requires many cores to run useful
simulations in reasonable time.  The CPU implementation bogs down when
the simulated lung contains many agents, 
but the GPU 
version 
always loops over the full space so it does not
suffer in this scenario."}

Our results and the developer feedback from both ADEPT and SIMCoV illustrate two scenarios in the software development cycle where EC-based optimization can help: rapid prototyping in the early development stage and advanced fine-tuning in the final development stage. In the prototyping stage, the developer can quickly implement a workable but less-optimized version of the software and let EC perform code optimization searches, identify potentially-interesting performance critical regions, and address those inefficiencies. In the late development stage, EC can be deployed after hand-tuning by experts to search for additional optimizations. 

A feature of our approach is that it does not require programmer domain knowledge for optimization. 
We acknowledge that EC-driven optimization does not necessarily preserve exact program semantics, which is both a strength and a limitation.  It is a strength because small changes in semantics can lead to large runtime reduction, often without sacrificing functionality. It is a limitation because test suites are used to evaluate fitness and
verify program behavior. With domain knowledge, developers can reason about the discovered optimizations, and either adopt them for better program performance, use them to improve the test suite, or use the insights to inspire related code enhancements, e.g., by introducing zero padding (Section~\ref{sec:result:boudary}). The results reported here for ADEPT did not require us to augment the test suite, an advantage of working with a deterministic program with an extensive test suite.  However, if there are mutations that improve performance but do not make sense to programmers, like the one that introduced an additional memory write into an unused code location (Section~\ref{sec:remaining}), it may make sense for the programmer to eliminate the edit or design new tests. 

GPUs are complex hardware with an equally complex programming environment. This is one reason why automated code optimization can be effective. Performant code can easily fail to provide expected performance, sending developers on a lengthy performance debugging journey. There is no golden rule for finding optimal performance on GPUs. For instance, higher concurrency does not guarantee higher performance, because in some cases using larger shared memory per block while minimizing occupancy may yield better throughput.  Similarly, as demonstrated in the case of ADEPT, 
using a faster method of inter-thread communication (register to register transfer) does not imply the best performance. 
EC can automate this search for counter-intuitive optimizations while exploring hundreds of times more code modifications than a human developer can reasonably consider. 
We expect that the results achieved for ADEPT may generalize to other bioinformatics kernels and programs that use dynamic programming. 

The final optimized sequence alignment program contains a large number of interacting edits, which is vastly more than what was reported by any earlier EC work for software (one or two edits is much more typical).  This could arise from several factors: basic properties of the LLVM-IR representation and mutation operators, properties of GPU architectures, opportunities presented by the particular algorithms, or the implementation choices made by the developer---an avenue for future work. In particular, more effective epistasis is discovered in ADEPT-V1 than in ADEPT-V0. The developer-optimized codes in ADEPT-V1 might provide more paths for epistasis to surface since those optimized codes seem to be more resilient to our mutation operators.  More generally, high-level languages are designed to help programmers express algorithms in a modular away that minimizes interactions between different parts of the code. So, it would not be surprising if their very structure works against epistasis. At the same time, the search space defined for a lower-level program representation like LLVM-IR is much larger than it is for source code, which would intuitively make search problems more challenging.  How these factors balance out, and how to measure them remains an open question. 

Regardless of their source, the fact that we found optimizations with such a high number of interacting edits shows how automated methods can discover complicated modifications to the target program, but it also presents challenges for deeper analysis. 
The approach used in the last step of our analysis involved exhaustively iterating through all the edit combinations. This will not scale well beyond the roughly twenty edits we considered. 

\section{Related Work}
\label{sec:related}

Beyond traditional compilers, the domain of automatic code optimization has three main branches: program synthesis~\cite{manna1980deductive, gulwani2011synthesis, alur2013syntax, Barthe2013Relational, Buchwald2018synthesizing}, superoptimization~\cite{schkufza2013stochastic, schkufza2014stochastic, conditionalCorrectSO, soundLoopSO}, and evolutionary computation~\cite{koza1994genetic, GP4ProgImprovement}. One key difference among the branches is the validation method. Program synthesis and superoptimization typically use a SAT/SMT solver~\cite{moura2008z3} to check the logical equivalence of program rewrites, while EC relies on testing sets to encode the intended program specification. The trade-off is that the SAT/SMT solver guarantees  program semantics but does not scale well, while test-based methods give up strict semantic equivalence but are more scalable. As a result,
most earlier work in this domain applies only to programs of a limited length, usually under 200 lines of codes.

Deep learning methods have recently been used to analyze programs as well, including neural-network based logical reasoning~\cite{evans2018can, paliwal2020graph} and SAT solvers~\cite{selsam2018learning, si2018learning}  but also for superoptimization~\cite{BunelDKTK17}. 
However, for optimizing parallel codes like GPU programs, EC may be more viable because logically reasoning about thread communications in a SAT solver requires deducing the entire parallel programming model in a logical form which is time-consuming and challenging.

EC is a popular approach for improving computer programs, e.g., to
automatically repair bugs~\cite{study4AutoProgRepair, fixes-suggest,genprog, Weimer:gp4patches,GP4AutoSWrepair}. Surprisingly, prior analysis~\cite{SWmutationRobust} showed that 20\% to 40\% of randomly generated program mutations (edits) have no observable functional effect (even when limited to only regions of the code that are actively tested), which suggested the possibility of using EC to optimize non-functional properties of software. As a result, EC has also been adopted to optimize software properties such as performance~\cite{GP4ProgImprovement} and energy cost~\cite{gp4energy, bruce15energy, bruce2018approximate, brownlee2021exploring}. 

Earlier EC work targeting GPU programs dates back to Sitthi-Amorn's work~\cite{gp4shader}, which began with a basic lighting algorithm and used EC to gradually modify the shader program into a
form that resembles an advanced algorithm proposed by domain experts. Later, Langdon et al. applied EC to a series of CUDA programs, ranging from compression methods~\cite{GP4cudaGzip} to RNA and DNA analysis~\cite{GP4cudaRNA, GP4cudaDNA}. Specifically,  BarraCUDA~\cite{klus2012barracuda}, a DNA sequence alignment program, was one of the target programs in the DNA analysis study~\cite{GP4cudaDNA}. However, their approach is different and less general than the one we used here. For example, the above works searched for parameter configurations outside the CUDA kernel such as the number of threads per thread block. The work manually parsed and transformed the CUDA kernel code into a custom-designed, line-based Backus Normal Form grammar as the code representation, where EC was applied. The performance improvements were attributed almost entirely to parameter tuning rather than modifying the kernel code. Orthogonal to the prior work, our approach finds performance optimization opportunities by transforming the implementation of functions. We instrument the modern LLVM compiler infrastructure to preprocess the CUDA program into LLVM-IR, a more general appoach that can be applied to any LLVM-IR program.

\section{Conclusion}

Optimizing GPU codes is a time-consuming process that requires deep knowledge in both the application domain and GPU architectures. This paper demonstrates the performance optimization potential of using GEVO on ADEPT, a GPU accelerated bioinformatics sequence alignment library, and SIMCoV, an agent-based COVID simulation of viral spread. We find improvements between 17\% - 29\% for ADEPT-V1, the expert-optimized version of ADEPT, and SIMCoV on various GPU platforms. Moreover, on ADEPT-V0, an earlier and less-optimized version, we find an incredible 30X improvement. This demonstrates the excellent potential of stochastic search methods such as GEVO to augment developer efforts to optimize GPU codes. 

Our analysis of the evolved optimizations 
points to multiple interdependent edits, which leads to performance improvements that are challenging for human experts to discover. This is one of the strengths of our method, which we believe can augment code optimization to find useful interdependencies beyond what is achievable by application developers. 
As GPU architectures are still rapidly evolving, the availability of an automated code optimization tool to discover hidden performance optimization opportunities will continue to be useful as an aid to the code development process. We expect such methods to play an increasingly important role in lifting developer burden from focusing disproportionately on optimization, especially in cross-domain developments such as bioinformatics and many important application domains.

\section*{Acknowledgements}
{
This work is supported in part by the National Science Foundation under grants CCF-1652132 and CCF-1618039, and used resources of the National Energy Research Scientific Computing Center (NERSC), a U.S. Department of Energy Office of Science User Facility located at Lawrence Berkeley National Laboratory, operated under Contract No. DE-AC02-05CH11231.
}

\bibliographystyle{IEEEtran}
\bibliography{ref}

\end{document}